\documentclass[prl,twocolumn]{revtex4}

\usepackage{graphicx}
\usepackage{dcolumn}
\usepackage{bm}

\usepackage{graphics}
\usepackage{graphicx}
\usepackage[dvips,usenames]{color}

\newcommand{\vs}{\quad}

\newcommand{\be}{\begin{equation}}
\newcommand{\ee}{\end{equation}}
\newcommand{\ba}{\begin{eqnarray}}
\newcommand{\ea}{\end{eqnarray}}

\begin{document}


\title{Almost exponential transverse spectra from power law spectra}
\author{Tam\'as S. Bir\'o}
\affiliation{KFKI, Research Institute for
Particle and Nuclear Physics, H-1525 Budapest, Hungary}
\author{Berndt M\"uller}
\affiliation{Physics Department, Duke University, Durham, NC-27708}
\date{printed \today}

\begin{abstract}
\vs
{\bf Abstract}
We point out that exponential shape of transverse spectra can be
obtained as the Fourier transform of the limiting distribution  
of randomly positioned partons with power law spectra given by pQCD,
which actually realize Tsallis distributions. 
Such spectra were used to obtain hadron yields by recombination in relativistic
heavy-ion collisions at RHIC energies.

\end{abstract}
\maketitle


\vs
In relativistic heavy-ion collisions
exponential transverse momentum distributions
has been observed at low to mid transverse momenta, $p_T \approx 1 - 4$ GeV
\cite{PHENIX,STAR,NA49}.
This fact was often interpreted as an indication that the source
of the particles can be described by a temperature.
This view climaxed in the ''thermal model'' which
assumes a situation well described by equilibrium thermodynamics
\cite{THERMAL}.

\vs
Regarding experimental data the exponential law is most prominent
in the transverse mass spectra which fit the form $\exp((m-m_T)/T)$
with $m_T=\sqrt{m^2 + p_T^2}$ at mid-rapidity at SPS and RHIC
energies quite well. A tendency of increasing slopes with
increasing hadron masses in these spectra is usually interpreted
as a sign of flow, on the basis of the relation $T = T_0 + m\langle u^2\rangle$
\cite{FLOW}.
A fit to pion, kaon and proton spectra at RHIC gives
$T_0 \approx 140$ MeV and $\langle u^2 \rangle \approx 0.5 - 0.6$
at freeze out\cite{RHIC-FLOW}.
A possible interpretation of this fit is a common flow pattern and a
{\em common temperature}, $T_0$, for all hadrons produced in the 
heavy-ion collision. This value, converted back to no-flow by using
the above formula for pions gives about $T \approx 210 $ MeV, 
which is somewhat above the color deconfinement
temperature predicted by lattice QCD calculations \cite{LQCD}.

\vs
Since there is no known hadronic process which could thermalize 
on such a short timescale at which this thermal state
must have developed, it is tempting to consider {\em quark level} 
mechanisms as reason for the observed exponential spectra.

\vs
Constituent quark and parton recombination models have been utilized
as a picture of hadron formation in relativistic heavy-ion collisions
in recent years \cite{ALCOR,RECOMB}. 
This idea roots in the constituent quark model of
hadrons originally combining mesons from two, baryons from three
quark-like partons. In order to obtain an exponential hadron spectrum
by this mechanism, one needs to start with parton spectra which
are already exponential in the intermediate $p_T$ regime of $2 - 4$
GeV. At higher transverse momenta these spectra go over into power laws,
as given by pQCD calculations. According to the respective regions of
dominance exponential and power law parton distributions are simply
added as a phenomenological effort to generate the desired experimental
hadron spectra.

\vs
In this letter we present a mechanism, by which power law
spectra of partons can be combined to give an exponential  spectrum.
For this process, one needs several partons to recombine to a
constituent quark in contrast to the final step of hadron
recombination where only two or three such dressed quarks
are used.
In the following we show that the exponential can be constructed as the
{\em limiting distribution} of cut power laws in the form
\be
  w(E) = a \left( 1 + \frac{E}{b} \right)^{-c}
\label{pQCDPART}
\ee
with parameters $a$, $b$ and $c$ fitted to pQCD calculation results
as in Refs.\cite{PARTON}. Here we modified somewhat the original
formula by using $E$, the parton energy, instead of the transverse
momentum $p_T$. At zero rapidity it is, however, equal to the transverse
mass $m_T$, and for large transverse momenta to $p_T$:
\be
 E = \cosh y \sqrt{p_T^2 + m^2}
\ee
For large transverse momenta $w(E)$ goes over into a power law spectrum: 
$w \approx ab^c \, |p_T|^{-c}$.

\begin{figure}[hb]
\begin{center}
\vspace*{-40mm}
\includegraphics[width=0.4\textwidth]{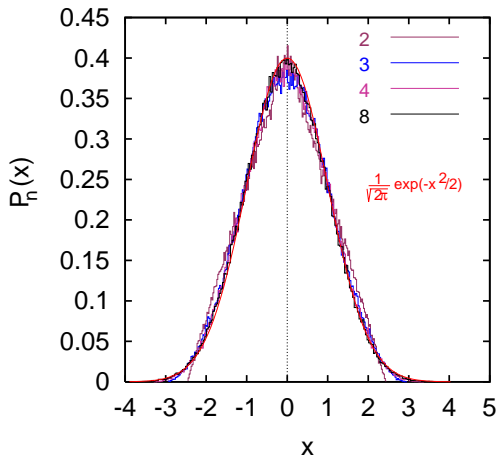}
\end{center}
\caption{\label{FigGauss}
 Comparison of histograms of $m = 200 000$ sums of $n$ uniform
  random deviates in $(-1,1)$ scaled with $\sqrt{3/n}$ and the
   limiting Gauss distribution $\frac{1}{\sqrt{2\pi}}\exp(-x^2/2)$.
   }
\end{figure}

\vs 
Before turning to the combination of power laws let us briefly recall
the familiar case of the central limit theorem \cite{CLIM}. 
Short tailed probability
distributions all lead to a Gaussian probability distribution for the
sum of random deviates in the limit of infinitely many independent
draws. It is easy to demonstrate this by considering uniform random deviates
$x_i$. Then $w(x_i)=0.5$ if $x_i$ is in the interval $[-1,1]$ and zero
otherwise. We seek the distribution $P_n(x)$ of the scaled sum,
\be
x = \sqrt{\frac{3}{n}} \sum_{i=1}^n x_i.
\ee
It is given by the n-fold integral
\be
 P_n(x) \: = \: 
   \prod_{i=1}^n \left( \frac{1}{2} \int_{-1}^{1}\!\! dx_i \right) \,
   \delta\left(x-\sqrt{\frac{3}{n}}\sum_{j=1}^n x_j\right).
\ee
In order to derive the Gaussian in the $n \rightarrow \infty$ limit
we use its Fourier transform,
\be
 \tilde{P}_n(k) \: = \: \int_{-\infty}^{+\infty} \!\! dx \,
  e^{ikx} \, P_n(x).
\ee
The $x$-integral is easily done, and the remainder
factorizes into $n$ equal contributions,
\be
 \tilde{P}_n(k) \: = \: \left(\frac{\sin(k\sqrt{3/n})}{k\sqrt{3/n}} \right)^n
\ee
In the large $n$ limit we arrive at
\be
 \tilde{P}_{\infty}(k) = \lim_{n\rightarrow\infty} \,
 \left( 1 - \frac{k^2}{2n} \right)^n \: = \: e^{-k^2/2}.
\ee
The inverse Fourier transformation results in the standard Gaussian
\be
 P_n(x) = \frac{1}{\sqrt{2\pi}} \, e^{-x^2/2}.
\ee 
Fig.\ref{FigGauss} demonstrates that this limit is approached very
quickly, already the sum of $n=3$  uniform random deviates
is distributed nearly Gaussian.


For the general case one uses an arbitrary distribution, $w(x_i)$
normalized to one, $\int w(x) dx = 1$. The central moments of
a distribution are given by higher derivatives of the logarithm
of the Fourier transform. For those of the $n$-fold sum we have
\be
 c_j(n) = \left(\frac{1}{i} \frac{\partial}{\partial k}\right)^j 
 	\left. \log \tilde{P}_n(k) \right|_{k=0}
\ee
Since $\log \tilde{P}_n(k) = n \log \tilde{w}(ka_n)$ when looking
for the distribution $P_n(x)$ of the scaled sum $x=a_n \sum_i x_i$,
we arrive at the following scaling law for the central moments
\be
 c_j(n) = n a_n^j c_j(1).
\ee
With a choice of $a_n \propto 1/\sqrt{n}$ one assures that
\be
 \lim_{n \rightarrow \infty} \,  c_j(n) = 0, \qquad {\rm for} \quad j>2.
\ee
Consequently the limiting distribution, $P_{\infty}(x)$, is Gaussian.

\vs
Now we consider the dressing of constituent quarks with gluons. 
One may  assume that the constituent energy and mass coalesces from
$n$ equal, partonic amounts -- a kind of additive parton model -- but
this is not necessary.
The total mass in this case would be $M = \sum E_i$, and each parton
would carry a fraction of the transverse momentum, 
$\lambda_i p_T = {\cal O}(p_T/n)$.
In the following we assume that the spectral shape of the 
constituent partons (recombining later to hadrons) is proportional
to the Fourier transform of the composite wave function.
Further we assume, that such a constituent quark (e.g. an up quark)
is composed of a bare up quark (the seed) and $n$ gluons (in general,
sea partons). Since these partons are almost massless, we consider
the center of energy as the proper coordinate for the constituent wave
function:
\be
 x = \frac{\sum E_ix_i}{\sum E_i} =  \sum \lambda_i x_i
\ee
with $\sum \lambda_i = 1$. By equal sharing of the total energy,
which is the case whenever the fusion happens at low relative
momenta, it is the arithmetic mean (with $1/n$ scaled sum) 
of random parton coordinates. 
Its distribution, $P_n(x)$, is interpreted in the present model as
the constituent quark wave function squared. Its Fourier
transform is proportional to the momentum spectrum and is obtained as
$n$-fold product of individual parton spectra, each carrying a
fraction of the total energy. For the sake of simplicity representing
all bare parton distribution $\tilde{w}$ by the same distribution,
we have, in the general case, $n$ contributions: 
\be
 \tilde{P}_n(k) \: = \: \prod_{i=1}^n \tilde{w}(\lambda_i k)
\ee
with $\sum_1^n \lambda_i = 1$.
Considering cut power law parton spectra as given by eq.(\ref{pQCDPART})
the $n$-fold spectrum at $y=0$ rapidity becomes
\be
 \tilde{P}_n(p_T) \: = \: \prod_{i=1}^n 
 	a_i \left(1+\frac{\lambda_i M_T}{b} \right)^{-c}.
\ee
Since all $\lambda_i$ are positive factors and their sum is normalized to one,
the limiting distribution for $n \rightarrow \infty$ 
enforces that all $\lambda_i \rightarrow 0$ and hence results in 
an {\em exponential} distribution, normalized to $\tilde{P}_{\infty}(0)=1$:
\be
 \tilde{P}_{\infty}(p_T) \: = \: 
   \exp \left(\frac{M-M_T}{b/c} \right).
\label{MAIN-RESULT}
\ee
\vs
This is the central result of the present letter.  It predicts a spectral slope
determined by pQCD as being $T = b/c$. From standard fits to the pQCD parton
spectra one gets $T \approx 170 - 210$ MeV, depending on the flavor under
consideration and details of the fit to the power law distribution from pQCD. 
This slope has nothing to do with the lattice QCD phase transition.

\newpage
\begin{figure}[htb]
\begin{center}
\vspace*{-30mm}
\includegraphics[width=0.48\textwidth]{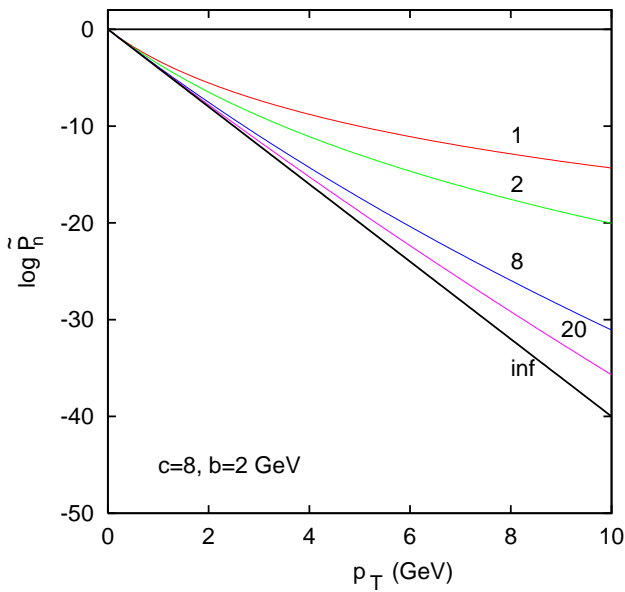}
\end{center}
\caption{\label{FigLIM}
  $\exp(-|p_T|c/b)$,
  the limiting distribution of cut power law Fourier transforms
  $(1+|p_T|/b)^{-c}$ and its finite $n$ approximations. 
  The value {$ n = 20 $} is characteristic for the pion.
}
\end{figure}

\vs
Our result can be understood by noticing that the cut power law
parton distribution of eq.(\ref{pQCDPART}) is the canonical
Tsallis distribution \cite{TSALIS} with a temperature $T = b/c$
and $q=1-1/c$. This distribution maximizes the Tsallis entropy
with the usual canonical constraint $\sum_i E_i w(E_i) = const$.
The down-scaled product of $n$ such distributions is still of the
Tsallis form at the same temperature $T=b/c$ but it belongs to
the parameter $q = 1 - 1/(cn)$. In the $n \rightarrow \infty$ 
limit $q\rightarrow 1$
and the usual canonical distribution is recovered with
the familiar exponential spectrum. By using the rather large values
for $c$ usually obtained by fitting pQCD results, namely $c=8$,
one obtains the following series: 
$q_1 \approx 0.875$, $q_2 \approx 0.94$, $q_3 \approx 0.96$, etc.


\vs
It is interesting to recall here another physical application
of the Tsallis distribution. Imperfectly thermalized systems,
e.g. a finite number of oscillators, also lead to a $q$ value
not equal to one. In the case of $(n+1)$ oscillators, however,
one arrives at a value bigger than one: $q=1+1/n$\cite{TSALIS}.
One obtains a distribution reminiscent of the discretized
path integral sum,
\be
 w_i = \frac{1}{Z} \left(1 - \frac{\varepsilon_i}{nT} \right)^n,
\ee
while primordial parton distributions belong to a $q$ value less than one,
\be
 w_i = \frac{1}{Z} \left(1 + \frac{\varepsilon_i}{nT} \right)^{-n}.
\ee
Of course, both distributions approach the Gibbs distribution
as $n\rightarrow\infty$.

\vs
We interpret the high power $c \approx 8-9$ in pQCD results as an
accidental value simulating an ''almost thermal'' distribution
with an effective $q \approx 0.82 - 0.89$ near to one. This explains
why the observed hadron spectra at low and intermediate $p_T$
appear nearly exponential in heavy-ion and many other collisions.
Parton recombination to hadrons amplifies this effect leading to
$q = 1 - 1/(cn)$ by  $n$-fold recombination. The Tsallis distribution
comes very close to the Gibbs distribution. As we shall argue in
the following, recombination processes are faster and more effective
with increasing phase space volume.
Consequently heavy-ion collisions are the best experiments to study
strongly interacting matter (QCD) as close to the canonical distribution
as possible. (It is a further interesting question whether the
primordial parton distribution is described by a smaller power $c$ in $e^+e^-$
collisions than in heavy-ion reactions. This fact could shed a new
light onto thermal model interpretation of hadron production in
$e^+e^-$ reactions \cite{Becattini}).
\vs
A remaining question is that how large $n$ has to be for $P_n$  to
look exponential. Fig.\ref{FigLIM} shows this behavior with 
the parameters $b=2$ GeV, $c=8$ (causing $T = 250$ MeV).
Here the convergence is slower than in the case of the sum of uniform
random deviates approaching a Gaussian.
The value $n=20$ would be characteristic for pion from additive
mass estimate $M_{\pi}/m_q \approx 21.5$ with $m_q=(m_u+m_d)/2$.


\vs
Another question is in which proportions the constituent partons, 
used for recombination to mesons and baryons, may consist of 
contributions from clusters of perturbative partons of different
complexity $n$.  Our suggestion
is, that one may not necessarily need to go far in $n$, i.e. to
use already exponential spectra for recombination, in order to
have an effective parton distribution close to the usually applied
one. In particular, and for the case of simplicity, we discuss
the $n=2$ case here: it improves already the original $n=1$
distribution a lot in the direction of the assumed exponential.

\begin{figure}[th]
\begin{center}
\vspace{-10mm}
\includegraphics[width=0.45\textwidth]{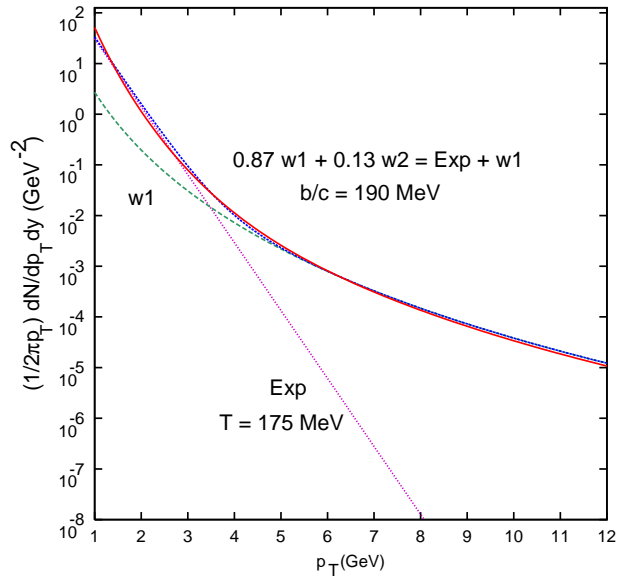}
\end{center}
\caption{\label{PYIELD}
 The thermal and power law parton yield components used by Fries. 
 et al.\cite{RECOMB},
 and the sum of the two contributions is compared to the 13-87\% 
 mixture of $n=1$ and $n=2$ parton clusters, using the power law
 distribution only.
}
\end{figure}

\vs
Fusion of a quark with a gluon and its radiation may be considered
as dynamical processes between $n=1$ and $n=2$ parton clusters.
In order to find an equilibrium ratio of these two type of partons
one inspects a rate equation. 
The detailed balance leads to the familiar proportionality applied
in recombination models:
\be
  f_2(E) = \frac{\gamma}{\Gamma} f_1^2(E/2).
  \label{BALANCE}
\ee
The effective spectrum we consider is given by
\be
 f(E) = A_1 w(E,b,c) + A_2 w^2(E, 2b, 2c).
\ee
Inspecting  the sum of thermal and original power law parton
distribution yields used for recombination by Fries. et al.
Ref.\cite{RECOMB}, one concludes that the ratio
$A_2/A_1 \approx 0.15$ gives already a surprisingly good approximation
in the relevant $p_T$ range from $b \approx 1.5$ GeV to $12$ GeV
(cf. Fig.\ref{PYIELD}). 
We note that in the original spectra parton energy loss
and a blue shift of the temperature of the thermal part
due to radial flow have been included. For our figure we 
used the same single parton yield, scaled down by a factor $0.87$ and
added its square at $p_T/2$ with weight $0.13$. The corresponding
Tsallis ''temperature'' is $b/c = 190$ MeV.

\vspace*{-30mm}


\newcommand{\qh}{{q\rightarrow h}}
\newcommand{\nqh}{{n_{\qh}}}
\vs
For realistic meson and baryon formation a further recombination should be
applied for each constituent of a hadron, possibly with different flavor,
in the final hadronization process. 
In a heavy-ion collision event many processes go on
in parallel. While some quarks are dressing up to become constituent quarks,
some other partons are fragmenting to lower $p_T$ hadrons
directly. The hadron $p_T$ spectra then are composed at
least from these two components, but maybe also of ''partially dressed''
quarks if a moderate $n$ in the production of their center of
energy from random positioned gluons appear. These may contribute
at intermediate $p_T$ to the hadron formation by recombination
as well; their yield is, however, reduced compared to the ones
behaving nearly exponentially.

\vs
The competition between fragmentation and recombination may in general
ease the entropy problem of the pure recombination models,
where finally one deals with fewer hadrons than partons. 

\vs
We analyze the entropy problem now for the final hadronization step.
As an approximation established by the above discussion, we consider
the recombination of exponential spectra from exponential precursors.
The entropy can be obtained using the Boltzmann formula (the $q=1$
case of the more general Tsallis entropy)
\be
S = \int (-f \ln f) \frac{d^3p}{(2\pi)^3} V
\ee
while the number of particles are given by
\be
N = \int \: f  \frac{d^3p}{(2\pi)^3} V.
\ee
The distribution $f(p)$ we consider is proportional to the Fourier
transform of the probabilities. Since they were normalized as $\tilde{P}_n(0)=1$,
here we have an unknown factor related to total numbers: $f=a\tilde{P}$.
For example for exponential spectra to combine, one gets
$f^n(p/n)=a^{n-1}*f(p)$. This leads to the familiar coalescence formula
$N_n = C_n N_1^n$ (with $C_n$ being proportional to $(VT^3)^{1-n}$) as well.
The $n$-cluster entropy becomes
\be
{S_n} =  (d-n \ln a) \, N_n,
\label{CLUSTER_ENT}
\ee
due to $\langle p \rangle/T=d$ when
considering $d$-dimensional momentum exponentials. At the same time the relative
number of $n$-clusters is related to the parameter 
$a$ as ${N_n}/{N_1} = a^{n-1}$.



\vs
The sum of all $n>1$-cluster entropies is then the entropy of
the recombined state, $S_H$, while $S_Q=S_1$ is that of the primordial state.
From the phase space integrals one gets
\be
S_H = \sum_{n>1} S_n = d \sum_{n>1} N_n - \ln a \sum_{n>1} (n N_n),
\ee
where the number of quarks redistributed in hadrons occurs as the
last term in the expression:
\be
\sum_{n>1} n N_n = N_1 = N_Q
\ee
is the recombination sum rule (counting $n$ quarks in an 
$n$-fold cluster). While this rule is quite strict
in the recombination of constituents, for dressing up quarks with gluons,
of course, such a constraint is not valid. Whenever this
sum rule applies, one obtains
\be
S_H = S_Q + d \left( N_H - \sum_{n>1} n N_n \right),
\ee
when expressing $ \ln a$ from the $n=1$ case of eq.(\ref{CLUSTER_ENT}).
Since the recombined sum goes over $n>1$ indices, and all $N_n$
are by definition positive, $\sum S_n < S_1$ is unavoidable
by the recombination of exponential spectra. 
With other words, after recombination there are fewer hadrons than
constituents built into hadrons.
The opening of several
different flavor channels may ease this problem, but cannot completely solve it.
The missing entropy is either generated by i) volume expansion
(i.e. the flow must be generated or accelerated by the recombination
process) or by ii) fragmentation, when a single parton produces
several hadrons (as it is known from $e^+e^-$ 2-jet events),
or by iii) the excitation of internal degrees of freedom of the hadrons.

\vs
Based on these principles all hadronic transverse momentum distributions
measured at RHIC can be worked out as outlined in this letter.
The previously assumed exponential parton spectra now are {\em derived} in the 
intermediate $p_T$ range by folding pQCD power law spectra. 
No assumption of a heat bath, a temperature
or thermal equilibrium is required. We present this scenario as an alternative
to, and at the same time a possible explanation for, the remarkable
phenomenological success of the thermal model.

\vs
We do not believe, however, that this result reduces the usefulness of
heavy-ion collisions in the search for quark matter. Asymmetric
jet quenching and the meson -- baryon scaling of the azimuthal
flow ($v_2$) indicate the presence of a collective state
with properties of a highly absorptive, perhaps colored, medium. The present
explanation of exponential transverse spectra from pQCD and random
statistics phenomenology offers an alternative to the 
ad hoc assumption of a thermal state. 
In fact, the independence of dressing gluons when combined to
a constituent parton is an ingredient of getting the
exponential $p_T$ distribution. This condition is probably satisfied only
in heavy-ion collisions.
Our model also calls attention to the fact that
''statistical'' is not always synonymous with equilibrium Gibbs thermodynamics.
Certain dynamical phenomena, like e.g. self organizing criticality
and -- what we have just discussed -- limiting parton distributions, 
belong to a more general framework of statistical physics.

\vs
The exponential $m_T$-spectra in fact approach at low $p_T$
a Gaussian, $\exp(-p_T^2/(2mT))$, the textbook case of a thermal
distribution. It is interesting to note that several recent
articles consider a Gauss distributed intrinsic parton $p_T$
in the generalized parton distribution functions (GPD-s) \cite{GPD}.
The Tsallis distribution predicts $\frac{1}{2}\langle p_T^2 \rangle =
m_T(b+m_T)$ leading to $\langle p_T^2 \rangle \approx 1$ GeV$^2$,
if using a constituent mass of $m_T=0.3$ GeV and the cut-off $b=1.5$ GeV,
in good agreement with recent theoretical analysis of high energy
$pp$ data.

\vs
We have seen, however, that the parameter $T$ in
this distribution is not related to a heat bath or another
source of statistical fluctuations. It is combined from two pQCD
parameters, from a cut-off $b$ and a power $c$ as being $T=b/c$.
The ''heating agent'' in a general sense here is the effective number
of independent dressing gluons per constituent, $n$, 
which should lead to a Tsallis distribution with $q=1-1/(cn)$
quite close to an exponential at low $p_T$.
If the transverse positions of the dressing partons are averaged 
as independent random
variables, the distribution of their center of energy will have a
Fourier spectrum, which is nearly exponential up to $p_T\approx 3$ GeV.

\vs
{\bf Acknowledgment} \quad
Discussions with J. Zim\'anyi and R. Fries are gratefully
acknowledged. Special thanks to C.Nonaka for making available
pQCD fit parameters and fragmentation function parameterizations.
T.S.B. thanks the hospitality of Duke University which was
expressed towards him during his stay in Durham.
This work has been supported in part by DOE grant
FG02-96ER-40945 and by a grant of the
Hungarian National Research Fund OTKA (T034269).


\end{document}